\documentclass[%
 reprint,
 amsmath,amssymb,
 aps,
]{revtex4-2}

\usepackage{graphicx}
\usepackage{dcolumn}
\usepackage{bm}
\usepackage{hyperref}
\usepackage{color}

\newcommand{\diffd}{\mathrm{d}} 

\begin{document}

\title{Testing the SZ-based tomographic approach to the thermal history of the universe with pressure-density cross-correlations: Insights from the \emph{Magneticum} simulation}

\author{Sam Young}
\email{syoung@mpa-garching.mpg.de}

\author{Eiichiro Komatsu}%
 \altaffiliation[Also at ]{Kavli Institute for the Physics and Mathematics of the Universe (Kavli IPMU, WPI), Todai Institutes for Advanced Study, The University of Tokyo, Kashiwa 277-8583, Japan.}
\affiliation{ 
Max Planck Institute for Astrophysics, Karl-Schwarzschild-Str. 1, 85748 Garching, Germany
}%

\author{Klaus Dolag}
\affiliation{University Observatory Munich, Scheinerstr. 1, 81679 Munich, Germany}
\affiliation{ 
Max Planck Institute for Astrophysics, Karl-Schwarzschild-Str. 1, 85748 Garching, Germany
}%

\date{\today}

\begin{abstract}
The thermal Sunyaev-Zeldovich effect contains information about the thermal history of the universe, observable in maps of the Compton $y$ parameter; however, it does not contain information about the redshift of the sources. Recent papers have utilized a tomographic approach, cross-correlating the Compton $y$ map with the locations of galaxies with known redshift, in order to deproject the signal along the line of sight. In this paper, we test the validity and accuracy of this tomographic approach to probe the thermal history of the universe. We use the state-of-the-art cosmological hydrodynamical simulation, \emph{Magneticum}, for which the thermal history of the universe is a known quantity. The key ingredient is the Compton-$y$-weighted halo bias, $b_y$, computed from the halo model.
We find that, at redshifts currently available, the method reproduces the correct mean thermal pressure (or the density-weighted mean temperature) to high accuracy, validating and confirming the results of previous papers. At higher redshifts ($z\gtrsim 2.5$), there is significant disagreement between $b_y$ from the halo model and the simulation.
\end{abstract}

\maketitle

\section{Introduction}
\label{sec:intro}

As cosmological structures form, the gravitational potential wells seeded by primordial density fluctuations become deeper \cite{peebles:1980}. As this process occurs, gravitational potential energy is converted into kinetic energy in an expended universe, following the Layzer-Irvine equation \cite{layzer:1963,irvine:1961,dmitriev/zeldovich:1963}.
A part of the kinetic energy is converted into thermal energy via the process of shock heating in the large-scale structure of the universe \cite{Cen:1998hc,Refregier:2000xz,Miniati:2000iu}, while the rest remains in the form of bulk/turbulent motion until it decays and thermalizes \cite{Shi:2014msa,Shi:2014lua}.
Therefore, measurements of the mean thermal energy of the cosmic gas content at different redshifts can be used to probe the growth of structure, as recently demonstrated in Refs.~\cite{Chiang:2020mzc,Chiang:2020ssj}.

To this end, the thermal Sunyaev-Zeldovch (SZ) effect \cite{Zeldovich:1969ff,Sunyaev:1972eq} can be used to probe the baryons in the universe \cite{Carlstrom:2002na,Kitayama:2014aaa,Mroczkowski:2018nrv}. As photons in the cosmic microwave background (CMB) are inverse-Compton scattered off of the free electrons in the ionized gas, they leave an imprint in the form of a spectral distortion in the CMB. The amplitude of the  SZ effect depends on the electron pressure integrated along the line of sight, which is parametrized by the Compton $y$ parameter \cite{Zeldovich:1969ff,Sunyaev:1972eq},
\begin{equation}
y(\hat{\phi}) = \frac{\sigma_\mathrm{T}}{m_\mathrm{e}c^2}\int \frac{\diffd \chi}{1+z}P_\mathrm{e}(\chi\hat{\phi}),
\end{equation}
where $\sigma_\mathrm{T}$ is the Thomson scattering cross section, $m_\mathrm{e}$ is the electron mass, $c$ is the speed of light, $\chi=\chi(z)$ is the comoving radial distance out to a given redshift $z$, and $P_\mathrm{e}$ is the electron pressure. The integral runs from zero up to the surface of last scattering, $z\simeq 1090$. The electron pressure $P_\mathrm{e}$ is related to the electron temperature $T_\mathrm{e}$ as $P_\mathrm{e} = n_\mathrm{e}k_\mathrm{B}T_\mathrm{e}$, where $n_\mathrm{e}$ and $k_\mathrm{B}$ are the (proper) electron number density and the Boltzmann constant, respectively.

The SZ effect, observable in maps of the Compton $y$ parameter, contains information about $P_\mathrm{e}$, but does not contain information about $z$ of the sources. To probe the growth history of structure using the SZ effect, we therefore require an external source of information about $z$ \cite{Zhang:2000wf,Shao:2009wz}. As the SZ signal is dominated by massive structure, the signal can be deprojected along the line of sight using clustering-based redshift inference \cite{Newman:2008mb,McQuinn:2013ib,Menard:2013aaa}. Following this method, an external sample of reference sources with known $z$ is taken, and cross-correlated with the Compton $y$ parameter as a function of $z$. This allows correlated intensities to be extracted tomographically \cite{Schmidt:2014jja,chiang/menard:2019,Chiang:2018miw}, as has been studied in recent papers for the SZ-galaxy cross-correlation with spectroscopic redshifts \cite{Vikram:2016dpo,Makiya:2018pda,Chiang:2020mzc}. See Refs.~\cite{Pandey:2019cml,Koukoufilippas:2019ilu,Yan:2021gfo} for the cross-correlation with photometric redshifts. 

The halo bias-weighted mean electron pressure, $\langle bP_\mathrm{e}\rangle$, is the direct observable of the SZ-galaxy cross-correlation function on large scales \cite{Vikram:2016dpo}. Here, $\langle ... \rangle$ denotes an ensemble average. To infer the mean  electron pressure, $\langle P_\mathrm{e}\rangle$, we need to know the Compton $y$-weighted halo bias, $b_y\equiv \langle bP_\mathrm{e}\rangle/\langle P_\mathrm{e}\rangle$. In Refs.~\cite{Chiang:2020mzc,Chiang:2020ssj}, the halo model developed in Refs.~\cite{Komatsu:1999ev,Komatsu:2002wc,Bolliet:2017lha,Makiya:2018pda,Makiya:2019lvm} was used to calculate $b_y$ and infer $\langle P_\mathrm{e}\rangle$ from the measured  $\langle bP_\mathrm{e}\rangle$.

How accurate is this approach? 
It is the aim of this paper to test the validity of the tomographic approach to the mean thermal history of the universe in Refs.~\cite{Chiang:2020mzc,Chiang:2020ssj}.
Specifically, we test this approach against the \emph{Magneticum} simulation \cite{Dolag:2015dta}. We use cross-correlations of the density and pressure to calculate the density-weighted mean temperature of baryonic gas, which is a known quantity in the simulation.

The rest of this paper is organized as follows. In Sec.~\ref{sec:simulation}, we describe the \emph{Magneticum} simulation.
We then present the comparison of the simulations and observations for $\langle bP_\mathrm{e}\rangle$ in Sec.~\ref{sec:bPe} and the density-weighted mean electron temperature $\bar T_\mathrm{e}$ in Sec.~\ref{sec:Te}, while we compare the halo model calculation and the simulation result for $b_y$ in Sec.~\ref{sec:by}.
We conclude in Sec.~\ref{sec:conclusion}.

\section{The \emph{Magneticum} simulation}
\label{sec:simulation}

The \emph{Magneticum} simulation is a set of state-of-the-art, cosmological hydrodynamical simulations of different cosmological volumes with different resolutions performed with an improved version of the smoothed-particle hydrodynamics (SPH) code \texttt{GADGET3} \cite{Springel:2005mi,Beck:2015qva}. They follow a standard $\Lambda$ Cold Dark Matter (CDM) cosmology with parameters close to the best-fitting values of the WMAP 7-year results \cite{Komatsu:2010fb} for a flat $\Lambda$CDM cosmology with the total matter density $\Omega_\mathrm{m} = 0.272$ (16.8\% baryons), the cosmological constant $\Omega_\Lambda = 0.728$, the Hubble constant $H_0 = 70.4~\mathrm{km~s^{-1}~Mpc^{-1}}$ ($h=0.704$), the index of the primordial power spectrum $n_\mathrm{s}=0.963$, and the overall normalisation of the power spectrum $\sigma_8 = 0.809$.

Here we describe the simulations briefly; for a more detailed description we refer to the previous work using these simulations \cite{Dolag:2015dta,Bocquet:2015pva,Gupta:2016yso,Soergel:2017ahb,Ragagnin:2019Mc,Castro:2021HMb}. The simulations follow a wide range of physical processes (see Refs.~\cite{Michaela:2013sia,teklu/etal:2015} for details) which are important for studying the formation of active galactic nuclei (AGN), galaxies, and galaxy groups and clusters. The simulation set covers a huge dynamical range following the same underlying treatment of the physical processes controlling galaxy formation, thereby allowing to reproduce the properties of the large-scale, inter-galactic and intra-cluster medium \cite{Dolag:2015dta,Gupta:2016yso,Remus:2017dns}, as well as the detailed properties of galaxies including morphological classifications and internal properties \cite{teklu/etal:2015,teklu/etal:2017,Remus:2016elq}. This also includes the distribution of different metal species within galaxies and galaxy clusters \cite{dolag/mevius/remus:2017} and the properties of the AGN population \cite{Michaela:2013sia,steinborn/etal:2016}. Especially the simulations well reproduce the observed pressure profiles of galaxy clusters \cite{Planck:2013pprof,Gupta:2016yso} and X-ray scaling relations \cite{Biffi:2013xobs}.

Here we focus on the largest box (`Box0') which follows the evolution of $2\times 4536^3$ particles in a large box of the comoving volume $\left( 2 688~h^{-1}~\mathrm{Mpc}\right)^3$ \cite{Bocquet:2015pva,Pollina:2016gsi,Soergel:2017ahb}, making it the largest cosmological hydrodynamical simulation performed to date.

\section{The halo bias-weighted mean electron pressure \texorpdfstring{$\langle bP_\mathrm{e}\rangle$}{<bPe>}}
\label{sec:bPe}

The tomographic technique provides direct constraints on the halo bias-weighted mean electron pressure, $\langle bP_\mathrm{e}\rangle$  \cite{Vikram:2016dpo}. This can be rewritten using the the large-scale, $y$-weighted halo clustering bias $b_y$ and the mean electron pressure $\langle P_\mathrm{e}\rangle$ as $\langle bP_\mathrm{e}  \rangle = b_y \langle P_\mathrm{e}\rangle$ \cite{Chiang:2020mzc}.
Making use of the \emph{Magneticum} simulation, knowledge of the density and pressure is readily available for different $z$, and instead of using the clustering redshift technique, we can perform an equivalent measurement using a cross-correlation of the density $\rho$ and thermal gas pressure $P_\mathrm{th}$:
\begin{equation}
\frac{\langle \delta_\rho P_\mathrm{th} \rangle(k)}{\langle \delta_\rho \delta_\rho \rangle(k)}\to \langle bP_\mathrm{th} \rangle =
b_y \langle P_\mathrm{th} \rangle,
\end{equation}
where $\delta_\rho \equiv (\rho - \bar{\rho})/\bar{\rho}$ and the bar denotes the mean value in the simulation. Assuming the gas is fully ionized, the electron pressure $P_\mathrm{e}$ is related to the total thermal gas pressure $P_\mathrm{th}$ as $P_\mathrm{th} = (8-5Y)/(4-2Y)P_\mathrm{e}$, where $Y=0.24$ is the primordial helium mass fraction (we assume a primordial mixture of elements and full ionization). Here, $\langle \delta_\rho P_\mathrm{th} \rangle(k)$ and $\langle \delta_\rho \delta_\rho \rangle(k)$ denote the density-pressure cross power spectrum and the density power spectrum, respectively, and $k$ is the wavenumber. The arrow indicates the large-scale limit, $k\to 0$.

Why do we call $\langle bP_\mathrm{th}\rangle$ the ``halo-bias weighted'' quantity, when we cross-correlate the pressure and density fields with no explicit reference to collapsed structures such as halos? The reason is that the thermal pressure is dominated by halos \cite{Refregier:2000xz,Komatsu:2002wc}; specifically, the density-pressure cross correlation is dominated by galaxy groups and clusters \cite{Chiang:2020mzc,Chiang:2020ssj}. Therefore, our estimator using the density fields yields the halo-bias weighted mean  pressure.

In order to obtain $\langle \delta_\rho P_\mathrm{th} \rangle(k)$ and $\langle \delta_\rho \delta_\rho \rangle(k)$, particles in the \emph{Magneticum} simulation are assigned to the nearest point in a $100^3$ grid spanning the simulation box. The coarse grain of the grid is acceptable since we are only interested in the large-scale limit of the cross-correlation. The \texttt{nbodykit} package \cite{Hand:2017pqn} is used to calculate the power spectra and cross-correlation of the resulting maps. 

\begin{figure}[t]
\centering
\includegraphics[width=0.45\textwidth]{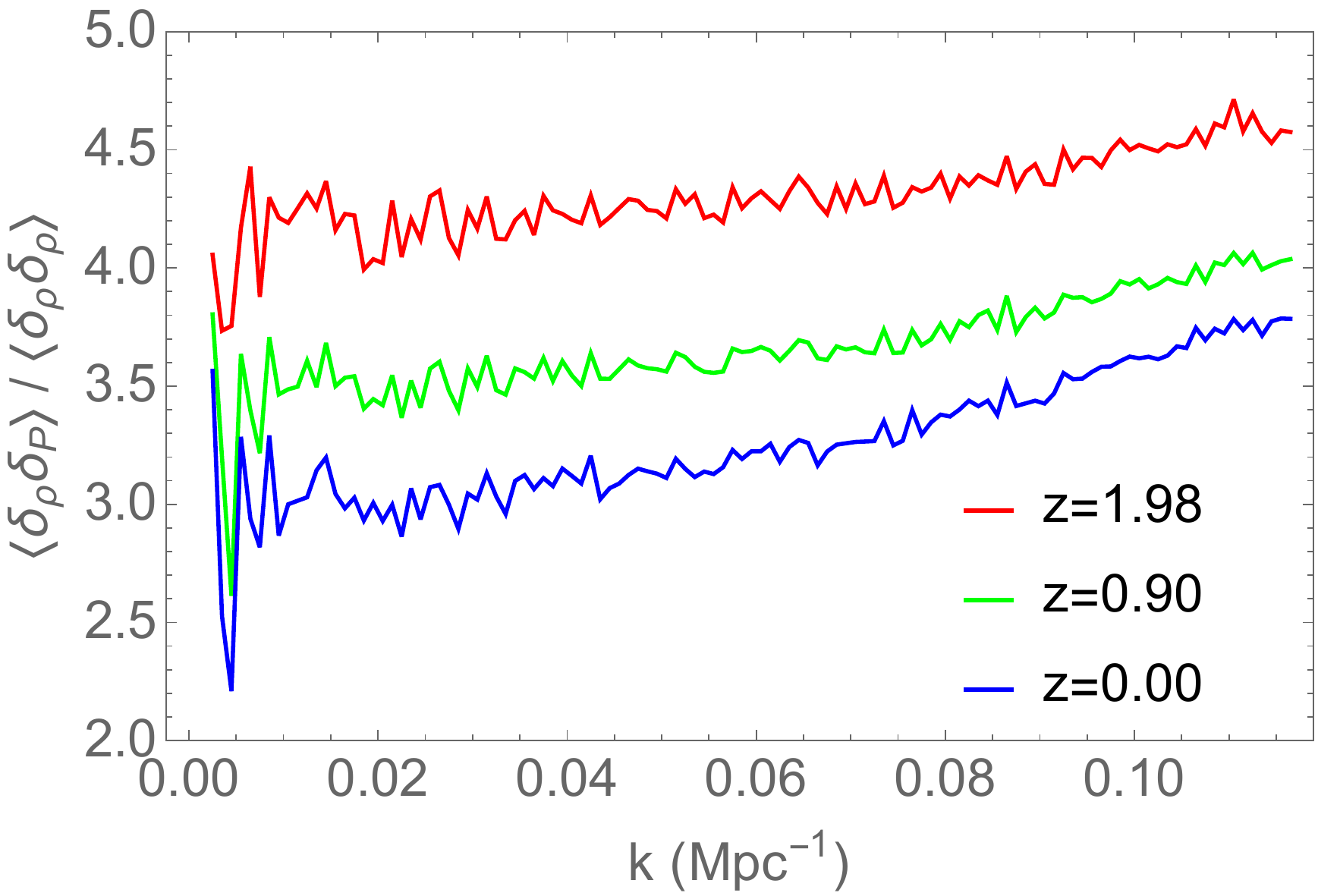}
\caption{The ratio of the density-pressure cross power spectrum and the density power spectrum, $\langle\delta_\rho \delta_P\rangle(k)/\langle\delta_\rho \delta_\rho\rangle(k)$ with $\delta_P\equiv (P_\mathrm{th}-\bar P_\mathrm{th})/\bar P_\mathrm{th}$, as measured in the \emph{Magneticum} simulation. We show the results for 3 different redshifts: $z=1.98$, $0.90$ and $0.00$ (from top to bottom). At large scales, the value approaches a constant value, $b_y$, to within scatter due to the finite box size.}
\label{fig:bias_k}
\end{figure}

In Fig.~\ref{fig:bias_k}, we show the ratio $\langle\delta_\rho \delta_P\rangle(k)/\langle\delta_\rho \delta_\rho\rangle(k)$ as a function of $k$, where $\delta_P\equiv (P_\mathrm{th}-\bar P_\mathrm{th})/\bar P_\mathrm{th}$. 
This quantity is equal to $b_y$ in the large-scale limit. 
We obtain the large-scale value by averaging the correlation functions over $k<0.03~\mathrm{Mpc}^{-1}$. This reduces the scatter in the correlation functions at small $k$ due to the finite box size. We have checked for convergence with respect to our choice of $k<0.03~\mathrm{Mpc}^{-1}$. The standard deviation in the correlation functions is used to derive error bars on the quantities derived using this method. We also checked that fitting the correlation functions to various forms has only a small effect on the calculated values for $b_y$ of approximately 2\%, which is subdominant to the uncertainty due to cosmic variance.

\begin{figure}[t]
\centering
\includegraphics[width=0.45\textwidth]{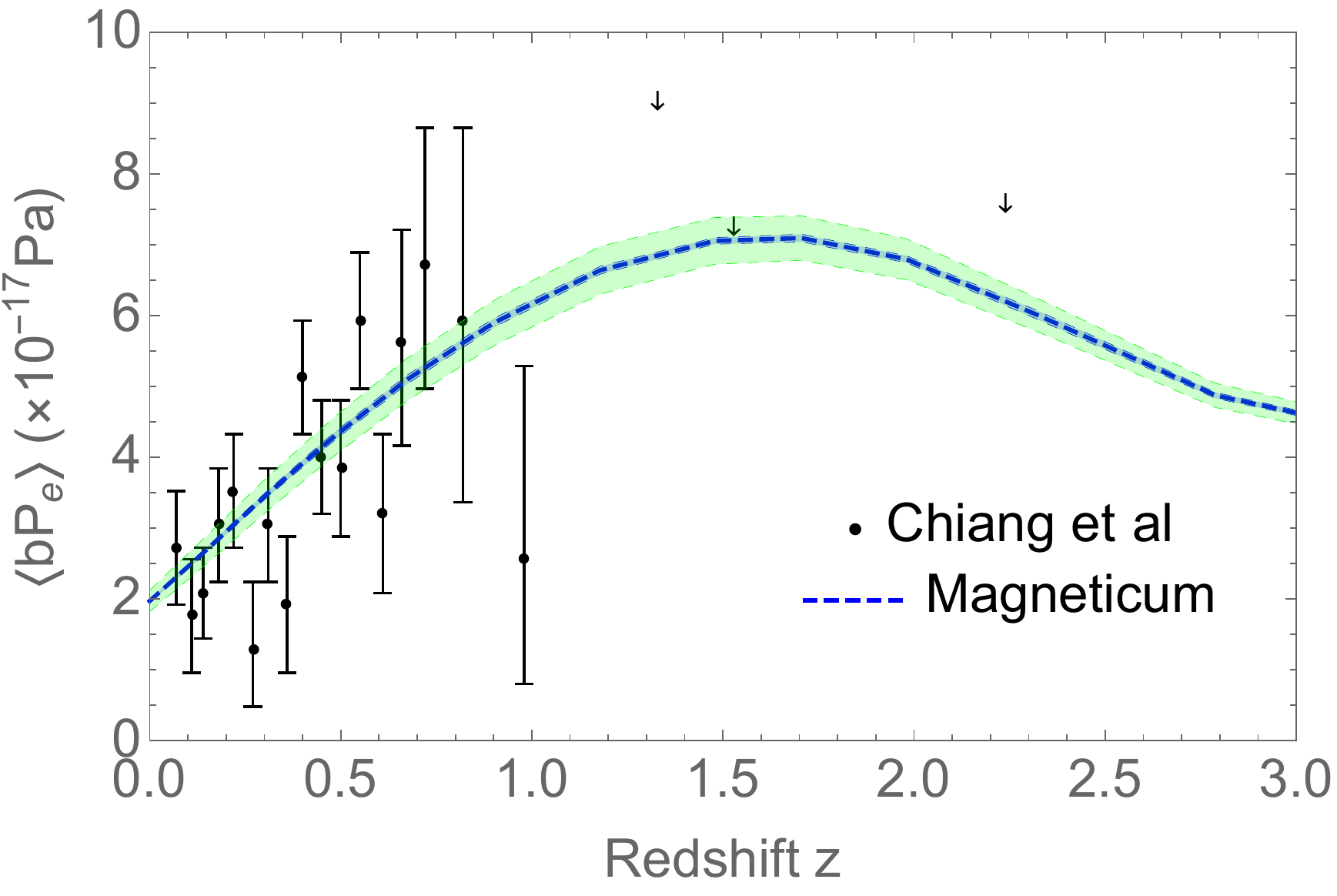}
\caption{The bias-weighted mean electron pressure $\langle bP_\mathrm{e} \rangle$ is shown as a function of $z$. The points are taken from the observed data given in Ref.~\cite{Chiang:2020mzc}, and the blue dashed line shows the value from the \emph{Magneticum} simulation. The (small) shaded blue region shows the $1\sigma$ uncertainty in the \emph{Magneticum} simulation due to the scatter in the correlation functions at small $k$, whilst the shaded green region shows the uncertainty due to cosmic variance.
}
\label{fig:bPe}
\end{figure}

We estimate the uncertainty due to cosmic variance by dividing the simulation box into 64 sub-regions and calculating the variance of $\langle b P_\mathrm{e}\rangle$ within those regions. 
In Fig.~\ref{fig:bPe}, we compare the values of $\langle b P_\mathrm{e}\rangle$  from the \emph{Magneticum} simulation and the observations \cite{Chiang:2020mzc}. They are in excellent agreement. To quantify how well the \emph{Magneticum} values fit the data, we consider an overall rescaling of the electron pressure $\langle bP_\mathrm{e} \rangle \rightarrow C \langle bP_\mathrm{e} \rangle$, and consider if this can give a better fit to the data as presented in \cite{Chiang:2020mzc}. A $\chi^2$ analysis, comparing the \emph{Magneticum} values with the data up to $z=0.98$ (above this value only upper bounds are available), gives a value $C=0.936\pm0.095$. This confirms that the measurements from the \emph{Magneticum} simulation are in agreement with observations - with an uncertainty at the $10\%$ level.

\section{The large-scale, \texorpdfstring{$y$}{y}-weighted halo clustering bias \texorpdfstring{$b_y$}{by}}
\label{sec:by}

While $\langle bP_\mathrm{e}\rangle$ is the direct observable of the SZ-based tomography, we need the knowledge of $b_y$ to obtain $\langle P_\mathrm{e}\rangle$ from the observed $\langle bP_\mathrm{e}\rangle$. In Refs.~\cite{Chiang:2020mzc,Chiang:2020ssj}, the halo model developed in Refs.~\cite{Komatsu:1999ev,Komatsu:2002wc,Bolliet:2017lha,Makiya:2018pda,Makiya:2019lvm}, which resulted in the \texttt{pysz} code \footnote{ \url{https://github.com/ryumakiya/pysz}}, was used to calculate $b_y$. How accurate is this approach?

The halo model calculation yields (see Appendix B of \cite{Chiang:2020mzc} for details)
\begin{equation}
b_y(z) = \frac{\int dM \frac{dn}{dM}M^{5/3+\alpha_p}b_h(M,z)}{\int dM\frac{dn}{dM}M^{5/3+\alpha_p}},
\label{eq:by}
\end{equation}
where $dn/dM$ and $b_h$ are the mass function and linear clustering bias of dark matter halos, respectively \cite{Tinker:2008ff,Tinker:2010my}. The physics is simple: the total pressure of gas in a halo $\int dVP_\mathrm{th}(M)$ is proportional to the halo mass $M$ times the virial temperature, the latter of which is proportional to $M^{2/3}$; thus, the virial relation gives $\int dVP_\mathrm{th}\propto M^{5/3}$. The X-ray observation of galaxy clusters shows a small empirical correction to this relation, $M^{5/3+\alpha_p}$ with $\alpha_p=0.12$ \cite{Arnaud:2009tt}.
We use the \texttt{pysz} code to calculate $b_y$ with the following parameters: $h=0.704$, $\Omega_\mathrm{b} h^2 = 0.02265$, $\Omega_\mathrm{c} h^2=0.11216$, $A_\mathrm{s} = 2.42\times 10^{-9}$, and $n_\mathrm{s}=0.963$.
The neutrino mass $m_\nu$ is set to zero because the \emph{Magneticum} simulation does not include massive neutrinos.

\begin{figure}[t]
\centering
\includegraphics[width=0.45\textwidth]{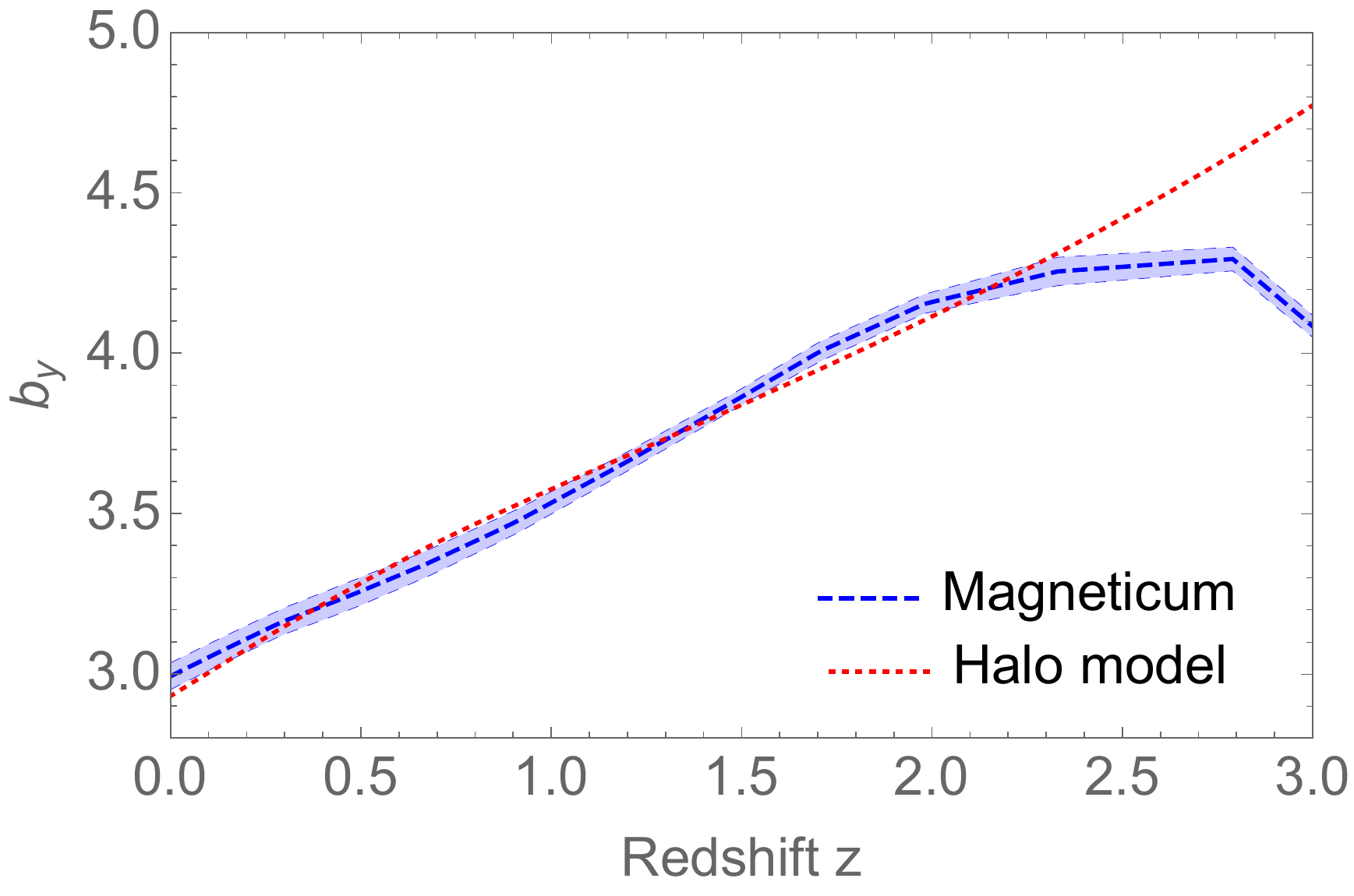}
\caption{The $y$-weighted halo bias $b_y$ is shown as a function of $z$. The red, dotted line is the prediction from the halo model, whilst the blue-line is that derived from the \emph{Magneticum} simulation. The shaded blue region represents the $1\sigma$ uncertainty in $b_y$ due to scatter in the large-scale correlation functions. We find excellent agreement between the two until $z\gtrsim 2.5$.}
\label{fig:bias}
\end{figure}

In Fig.~\ref{fig:bias}, we show the comparison of $b_y$ calculated with the halo model and from \emph{Magneticum}. We find that for low $z$, there is excellent agreement between the two calculations - well within the uncertainties. However, at large $z\gtrsim 2.5$, there is significant disagreement, with $b_y$ measured in \emph{Magneticum} being significantly lower than predicted by the halo model. This suggests that the measurements made in Ref.~\cite{Chiang:2020mzc} of the mean electron temperature (see Sec.~\ref{sec:Te}) are reliable up to a redshift $z\lesssim 2.5$, but implies uncertainty in the upper bounds which were derived at $z\gtrsim 2.5$. 

The reason for this disagreement is not clear, but it is plausible that the assumption (made for the halo model) that pressure is dominated by the virialized structures and the contribution from supernova and AGN feedback are subdominant compared to the thermal pressure of virialized gas may be violated at such a high $z$. 
There may be an additional term arising from a response of internal properties of galaxy groups and clusters to the large-scale overdensity \cite{Afshordi:2003xu,Voivodic:2020bec}, which we do not include in Eq.~(\ref{eq:by}). As Eq.~(\ref{eq:by}) agrees with $b_y$ from the \emph{Magneticum} simulation, which should include all the relevant effects, precisely at $z\lesssim 2.5$, these contributions are small in this redshift range. Further analysis of this disagreement at $z\gtrsim 2.5$ is left for future work.

To quantify how well the \emph{Magneticum} values fit the halo model prediction, we again consider an overall rescaling factor to the halo bias calculated with the halo model, $b_{y} \rightarrow A b_{y}$, and consider whether this can provide a better fit to the the simulation. Performing a $\chi^2$ analysis (and excluding values from $z>2.5$) gives a value $A=1.0027\pm 0.0043$, indicating that the halo model prediction is able to predict the values from \emph{Magneticum} to better than $1\%$ accuracy.

\section{The density-weighted mean electron temperature \texorpdfstring{$\bar T_\mathrm{e}$}{Te}}
\label{sec:Te}

The mean electron pressure $\langle P_\mathrm{e}\rangle$ is related to the density-weighted mean temperature of electrons in the universe, defined as $\bar{T}_\mathrm{e} \equiv {\langle n_\mathrm{e}T_\mathrm{e}\rangle}/{\langle n_\mathrm{e}\rangle}$ \cite{Cen:1998hc,Refregier:2000xz}. Specifically, using $\langle bP_\mathrm{e}\rangle$ and $b_y$, we find
\begin{equation}
\bar{T}_\mathrm{e} = \frac{2 m_\mathrm{H}}{\rho_\mathrm{c}\Omega_\mathrm{b}k_\mathrm{B}(2-Y)(1+z)^3}\frac{1}{b_y}\langle b P_\mathrm{e} \rangle,
\end{equation}
where $\rho_\mathrm{c} = 3H_0^2/(8\pi G)$ is the critical density of the universe at $z=0$ and $m_\mathrm{H}$ is the hydrogen mass.

The temperature data can also be read directly from the output files of the \emph{Magneticum} simulation, and the uncertainty in the mean temperature (and pressure) due to cosmic variance is estimated by dividing the simulation box into 64 sub-regions and calculating the variance of the mean temperature within those regions. We assume throughout that the electron temperature and the total temperature are equal, although there is likely to be some difference between the two, as discussed in more detail in Refs.~\cite{Akahori:2009yj,Avestruz:2014dea}.

\begin{figure}[t]
\centering
\includegraphics[width=0.45\textwidth]{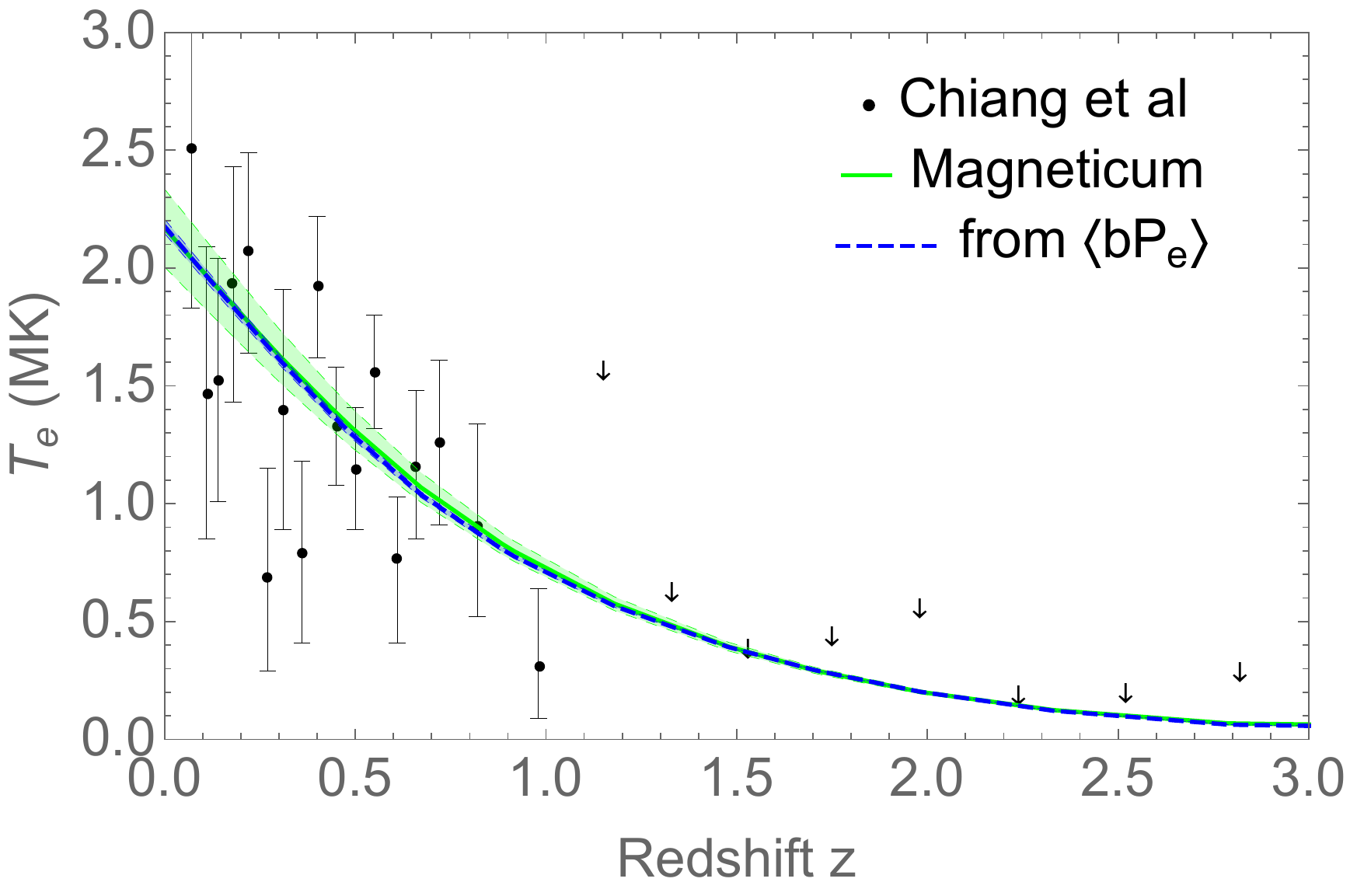}
\caption{The density-weighted mean electron temperature $\bar{T}_\mathrm{e}$ (in units of million K) is shown as a function of $z$. The points are taken from the observed data given in Ref.~\cite{Chiang:2020mzc}.
The solid green line shows the temperature read directly from the \emph{Magneticum} simulation, while the blue dashed line shows that calculated from the pressure data in the \emph{Magneticum} simulation. The blue and green regions show the $1\sigma$ confidence intervals for their respective data.
}
\label{fig:temp}
\end{figure}

For a direct comparison to the measurements of Ref.~\cite{Chiang:2020mzc}, we here use the value of $b_y$ calculated using the halo model, rather than the measured value from the \emph{Magneticum} simulation. In Fig.~\ref{fig:temp}, we compare  
$\bar T_\mathrm{e}$ from the data and the simulation
as a function of $z$. As expected from the excellent agreement for $\langle bP_\mathrm{e}\rangle$ (Fig.~\ref{fig:bPe}) and $b_y$ (Fig.~\ref{fig:bias}), we find excellent agreement for $\bar T_\mathrm{e}$. This completes the validation of the methodology for obtaining $\langle T_\mathrm{e}\rangle$, i.e., the thermal history of the universe, developed in Refs.~\cite{Chiang:2020mzc,Chiang:2020ssj}.

\section{Conclusion}
\label{sec:conclusion}

This study aimed to answer two principle questions:
\begin{enumerate}

\item The bias-weighted mean electron pressure $\langle bP_\mathrm{e} \rangle$ is observable from cosmological surveys. Does this quantity measured from the \emph{Magneticum} simulation agree with the data given in Ref.~\cite{Chiang:2020mzc}?

\item In Ref.~\cite{Chiang:2020mzc}, the density-weighted mean electron temperature $\bar T_\mathrm{e}$ is derived by dividing $\langle bP_\mathrm{e} \rangle$ by $b_y$ calculated from the halo model. Is this method accurate?

\end{enumerate}

The results presented here show that the answer to both questions is: yes. The values of $\langle bP_\mathrm{e} \rangle$ (as well as $\bar T_\mathrm{e}$) from \emph{Magneticum} agree well with the data - with the best-fitting value to rescale the overall pressure given by $C=0.936\pm0.095$, consistent with unity to within the uncertainty. Likewise, the halo model prediction for $b_y$ matches very well to that measured in \emph{Magneticum}, with the best-fitting rescaling value given by $A=1.0027\pm 0.0043$.

The results presented here confirm the validity of the tomographic method in order to determine the thermal history of gas in the universe, although highlight a need for further study of the halo model calculation at high redshifts, $z\gtrsim 2.5$, which are yet to be probed by observations. 

We also suggest that the measured mean temperature can be used as a simple ``thermometer test'' of the baryonic physics of simulations to confirm that they are capable of accurately reproducing observed values, which can be performed for a host of state-of-the-art cosmological hydrodynamical simulations \cite{vogelsberger/etal:2014,dubois/etal:2014,lebrun/etal:2014,schaye/etal:2015,mccarthy/etal:2017,pillepich/etal:2018}. The \emph{Magneticum} simulation has passed this test. How about yours?

\begin{acknowledgments}
EK thanks Y.-K. Chiang, R. Makiya and B. M\'enard for the collaboration on Refs.~\cite{Chiang:2020mzc,Chiang:2020ssj}, which motivated this work. SY thanks R. Makiya for his help with the \texttt{pysz} code \cite{Makiya:2018pda,Makiya:2019lvm} and was supported during the research by a Humboldt Research Fellowship.
This work was supported in part by the Deutsche Forschungsgemeinschaft (DFG, German Research Foundation) under Germany's Excellence Strategy - EXC-2094 - 390783311. KD acknowledges support for the COMPLEX project from the European Research Council (ERC) under the European Union’s Horizon 2020 research and innovation program grant agreement ERC-2019-AdG 860744. The Kavli IPMU is supported by World Premier International Research Center Initiative (WPI), MEXT, Japan. The calculations were carried out at the Leibniz Supercomputer  Center (LRZ) under the project pr83li. We are especially grateful for the support by M. Petkova through the Computational Center for Particle and Astrophysics (C2PAP) and the support by N. Hammer at LRZ when carrying out the Box0 simulation with the Extreme Scale-Out Phase on the new SuperMUC Haswell extension system.
\end{acknowledgments}

\bibliography{bibfile}

\end{document}